\documentclass[aps,11pt]{revtex4}
\usepackage[colorlinks=true,linkcolor=blue,urlcolor=blue,filecolor=black,citecolor=red,pdfstartview=FitV,pdftitle={},pdfsubject={},
pdfkeywords={},pdfpagemode=None,bookmarksopen=true]{hyperref}
\usepackage{graphicx}
\usepackage{epstopdf}%
\usepackage{amsmath}
\usepackage{amsfonts}
\usepackage{amssymb}
\usepackage{color}%
\usepackage{dcolumn}
\usepackage{slashed}
\usepackage{amssymb,ulem}
\usepackage{float}
\usepackage{amsthm,amsmath,amssymb}
\usepackage{mathrsfs}
\usepackage{subfigure}
\usepackage{wrapfig}
\usepackage{indentfirst}
\usepackage{multirow}
\usepackage{appendix}
\makeatletter

\newcommand{\Rmnum}[1]{\expandafter\@slowromancap\romannumeral #1@}

\makeatother

\begin{document}
\title{Black holes in massive Einstein-dilaton gravity}
\author{Bo Liu$^{1,2}$\footnote{fenxiao2001@163.com;}, Rui-Hong Yue$^{3}$\footnote{rhyue@yzu.edu.cn;}, De-Cheng Zou$^{4,3}$\footnote{Corresponding author: dczou@jxnu.edu.cn}, Lina Zhang$^{5}$\footnote{linazhang@hunnu.edu.cn}, \\ Zhan-Ying Yang$^{1}$\footnote{zyyang@nwu.edu.cn;} and Qiyuan Pan$^{5}$\footnote{panqiyuan@hunnu.edu.cn;}  }

\affiliation{$^{1}$School of Physics, Northwest University, Xi'an, 710127, China\\
$^{2}$School of Arts and Sciences, Shaanxi University of Science and Technology, Xi'an, 710021, China\\
$^{3}$Center for Gravitation and Cosmology, College of Physical Science and Technology, Yangzhou University, Yangzhou 225009, China\\
$^{4}$College of Physics and Communication Electronics, Jiangxi Normal  University, Nanchang 330022, China\\
$^{5}$ Key Laboratory of Low Dimensional Quantum Structures and Quantum Control of Ministry of Education, Synergetic Innovation Center for Quantum Effects and Applications, and Department of Physics, Hunan Normal University, Changsha, Hunan 410081, China}

\date{\today}

\begin{abstract}
\indent

In this paper, we focus on massive Einstein-dilaton gravity including the coupling of dilaton scalar field
to massive graviton terms, and then derive static and spherically symmetric solutions of dilatonic black
holes in four dimensional spacetime. We find that the dilatonic black hole could possess two horizons
(event and cosmological), extreme (Nariai) and naked singularity for the suitably fixed parameters.
In addition, we investigate thermodynamic properties of these dilatonic black holes, and check the
corresponding first law of black hole thermodynamics. Extending to the massive Einstein-dilaton gravity in
high dimensions, we further obtain the dilatonic black hole solutions in ($d+1$) dimensional spacetime. 
\end{abstract}

\maketitle

\section{Introduction}
\label{1s}

Despite many successes agreement with observations, the Einstein's general relativity(GR) might be searched for alternatives
due to  the cosmological constant problem \cite{weinberg:1989}, and the origin of acceleration of our universe based on the supernova data \cite{Riess:1998,Perlmutter:1999} and cosmic microwave background
(CMB) radiation \cite{Planck:2015fie,WMAP:2003elm}. An alternative theory of GR is dilaton gravity, coming from the low energy limit of
string theory, in which Einstein's gravity is recovered along with a scalar
dilaton field  by nonminimal coupling to  other fields such as axion and gauge fields \cite{Green:1987sw}.
The presence of dilaton field is necessary and plays an essential role in string theory
if one couples the gravity to other gauge fields. Until now, many attempts have been made to investigate the dilaton gravity theory.
For instance, Refs.~\cite{Mignemi:1991wa}-\cite{Sheykhi:2007wg} discovered that the dilaton field changes the causal structure of the black hole and leads to the curvature singularities at finite radii. The dilaton potential can be regarded as the generalization of the cosmological constant,
and also change the asymptotic behavior of the solutions.
Combined three Liouville-type dilaton potentials,
the black hole solutions of dilaton gravity in the background of (A)dS spaces were investigated in Refs.\cite{Gao:2004tu,Gao:2004tv}.
In addition, the scalar-tensor type generalizations of general relativity have been also investigated by containing various kinds of curvature corrections to the usual Einstein-Hilbert Lagrangian coupled to
the dilaton scalar field \cite{Berti:2015itd,Pani:2011xm,Yunes:2011we}.
A particular model called the  Einstein-dilaton-Gauss-Bonnet (EdGB) gravity was extensively studied in Refs.~\cite{Kanti:1995vq,Torii:1996yi}.
It found that the scalar dilaton is a secondary hair because the dilaton charge is expressed in terms of the black hole mass.
Later, the black holes in various dimensions~\cite{Guo:2008hf,Ohta:2009tb,Ohta:2009pe}.
 rotating black holes \cite{Kleihaus:2011tg,Maselli:2015tta},
wormholes \cite{Kanti:2011jz},
and rapidly rotating neutron stars \cite{Kleihaus:2016dui} were investigated in EdGB gravity.

From the perspective of modern particle physics \cite{Weinberg:1965pg,Boulware:1975cg},
gravity field can be treated as a unique theory of a spin-2 graviton. Massive gravity is a straight forward and natural modification by simply giving a mass to the graviton, dating back to 1939 when Fierz and Pauli \cite{Fierz:1939}
constructed a linear theory of massive gravity, which is always plagued with the Boulware-Deser ghost in non-linear level \cite{Boulware:1972if,Boulware:1972fr}. Notice that the authors of \cite{Rham:2011tl} constructed a theory where the
Boulware-Deser ghost was eliminated by introducing
higher order interaction terms into the Lagrangian.
Then, the ghost-free massive theory known as dGRT massive gravity was discussed in Refs.~\cite{Hinterbichler:2012, Rham:2014mg}.
In dGRT massive gravity, a class of (charged) black hole solutions \cite{Vegh:2013}-\cite{Ghosh:2015cva} and their corresponding thermodynamics \cite{Cai:2015tb}-\cite{Zou:2017juz}
in asymptotically AdS spacetime were investigated, and the coefficients in the potential associated with the graviton mass were shown to play the same roles as as the charge in thermodynamic phase space. Other solutions of black holes were also studied in massive gravity\cite{Babichev:2014rb}-\cite{Hendi:2018gb}.
Recently, the so-called Quasi-dilaton massive gravity has been also investigated in Refs. \cite{DAmico:2012hia}-\cite{Akbarieh:2021vhv},
which are scalar extended dRGT massive gravity with a shift symmetry.
Inspired by these, we would like to extend the study by considering the nonminimal coupling of dilaton field to graviton, and derive analytically solutions of dilatonic black holes in massive dilaton gravity.

The paper is organized as follows. In Sec.~\ref{2s}, we will present the static and spherically symmetric  black hole solutions in four dimensional massive Einstein-dilaton  gravity, and investigate the solution structures of dilatonic black holes.  In Sec.~\ref{3s}, we will discuss the thermodynamic properties of these black holes. Considering the massive Einstein-dilaton gravity in high dimensional spacetime, we will derive the ($d+1$) dimensional solutions of black hole in Sec.~\ref{4s}. Finally, we close the paper with discussions and conclusions in Sec.~\ref{5s}.

\section{BLACK HOLE IN MASSIVE EINSTEIN-DILATON GRAVITY}
\label{2s}

The action for massive gravity with a nonminimal coupling of dilaton field $\varphi$ in four dimensional spacetime is given by
\begin{equation}\label{eq4:action}
  I=\frac{1}{16 \pi} \int d^{4} x \sqrt{-g}
  \Big[\mathcal {R}  -2(\nabla \varphi)^2-V(\varphi)
  + m_0^2 \sum_{i=1}^{4} c_i e^{-2\beta_i \varphi} \mathcal {U}_i(g,h)\Big],
\end{equation}
where $\varphi=\varphi(r)$ is the dilaton scalar field, and $V(\varphi)$ is a potential for $\varphi$.
The last term in the action denotes general form of nonminimal coupling between the scalar field and massive graviton with coupling constants $\beta_i$. Here $m_0$ is the mass of graviton, and $c_i$ are the number of dimensionless coupling coefficients.
Moreover, $\mathcal{U}_i$ are symmetric polynomials of the eigenvalues of the ~$4\times 4$\ matrix
$K^{\mu}_{~\nu}=\sqrt{g^{\mu\alpha}h_{\alpha\nu}}$ in which $h$ is a fixed rank-2 symmetric tensor, satisfying the following recursion relation~\cite{Hinterbichler:2012}
\begin{eqnarray}\label{eq4:u1234}
  \mathcal{U}_1& =& [K]=K^{\mu}_{~\mu}, \nonumber\\
  \mathcal{U}_2 &=& [K]^2 - [K^2], \nonumber\\
  \mathcal{U}_3 &=& [K]^3 - 3[K][K^2] + 2[K^3],\\
  \mathcal{U}_4 &=& [K]^4 - 6[K^2][K]^2 + 8[K^3][K] + 3[K^2]^2 - 6[K^4].\nonumber
  \end{eqnarray}
  
Varying the action with respect to the field variables $g_{\mu\nu}$ and $\varphi$, the equations of motion are obtained as
\begin{eqnarray}
  G_{\mu\nu}&=& \mathcal {R}_{\mu\nu}-\frac{1}{2} \mathcal {R}g_{\mu\nu}= 2\partial_{\mu}\varphi\partial_{\nu}\varphi
                  -\frac{1}{2}\big[V+2 (\nabla\varphi)^2\big]g_{\mu\nu}
                 +m_0^2 \chi_{\mu\nu}, \label{eq4:eist} \\
  \nabla^2\varphi  &=& \frac{1}{4}\big[\frac{\partial V}{\partial \varphi}
                  -m_0^2 \sum_{i=1}^{4} \frac{\partial \tilde{c}_i}{\partial \varphi} \mathcal {U}_i
                  \big],\label{eq4:dilaton}
\end{eqnarray}
where
\begin{eqnarray}
  \tilde{c}_i&=&c_i e^{-2\beta_i \varphi},\\ \nonumber
  \chi_{\mu\nu}&=& \frac{\tilde{c}_1}{2}(\mathcal{U}_1 g_{\mu\nu}-K_{\mu\nu})
  +\frac{\tilde{c}_2}{2}(\mathcal{U}_2 g_{\mu\nu}-2\mathcal{U}_1 K_{\mu\nu}
  +2K^2_{\mu\nu})\nonumber\\
  &+& \frac{\tilde{c}_3}{2}(\mathcal{U}_3 g_{\mu\nu}-3\mathcal{U}_2 K_{\mu\nu}
  +6\mathcal{U}_1 K^2_{\mu\nu}-6K^3_{\mu\nu})\nonumber \\
  &+& \frac{\tilde{c}_4}{2}(\mathcal{U}_4 g_{\mu\nu}-4\mathcal{U}_3 K_{\mu\nu}
  +12\mathcal{U}_2 K^2_{\mu\nu}-24\mathcal{U}_1 K^3_{\mu\nu}
  +24K^4_{\mu\nu})\label{eq4:chi}.
\end{eqnarray}

Now we introduce the static and spherical symmetry metric ansatz
\begin{eqnarray}\label{eq4:metric}
  ds^2 = -f(r) dt^2 +f^{-1}(r) dr^2+ r^2R^2(r) d\Omega^2,
\end{eqnarray}
in which $f(r)$ and $R(r)$ are functions of $r$ and $d\Omega^2=d\theta^2+\sin^2\theta d\phi^2$\ is the line element for two dimensional spherical  subspace  with constant curvature.

Since the fiducial metric $h_{\mu\nu}$ in the action \eqref{eq4:action} plays the role of a Lagrange multiplier to eliminate the BD ghost~\cite{Vegh:2013}, one can choose an appropriate form to  simplify the calculation. The authors of Ref.\cite{Vegh:2013} pointed out that distinguished from the dynamical physical metric $g_{\mu\nu}$, the reference metric $h_{\mu\nu}$ is usually fixed and assumed to be non-dynamical in the massive theory. In this work, we will follow \cite{Cai:2015tb,Xu:2015rfa} by choosing the fiducial metric to be
\begin{eqnarray}\label{eq4:ref metr}
h_{\mu\nu}=diag(0, 0, c_0^2, c_0^2\sin^2\theta),
\end{eqnarray}
where $c_0$ is a positive parameter and we set $c_0=1$ for convenience in whole paper.

From the ansatz (\ref{eq4:ref metr}), the interaction potential in Eq.~(\ref{eq4:u1234}) changes into
\begin{eqnarray}\label{u1:u2}
\mathcal{U}_1=\frac{2}{R r},\ \ \
\mathcal{U}_2=\frac{2}{R^2 r^{2}},\ \ \
\mathcal{U}_3=\mathcal{U}_4=0.
\end{eqnarray}
Then, $\chi^{\mu}_{~\nu}$ in Eq.(\ref{eq4:chi}) becomes
\begin{eqnarray}\label{eq4:chi:1}
\chi^1_{~1} = \chi^2_{~2} = \frac{c_1 r R e^{-2\beta_1 \phi }   + c_2 c_0^2 e^{-2\beta_2 \phi }}{(r R)^2 },\ \ \
\chi^3_{~3} = \chi^4_{~4} = \frac{c_1 e^{-2\beta_1 \phi }}{2 r R }
\end{eqnarray}
and the corresponding components of equation of motion (\ref{eq4:eist}) can be simplified to
\begin{eqnarray}
G^1_{~1}&=& \frac{1}{(r R)^2}[r R (r R)'f'+2r R (r R)''f+(r R)'^2 f-1] =-\frac{1}{2}V(\varphi)-f \varphi'^2+m_0^2 \chi^1_{~1}
\label{eq4:inseq:11},\\
  G^2_{~2} &=& \frac{1}{(r R)^2}[r R (r R)'f'+(r R)'^2 f -1 ]=-\frac{1}{2}V(\varphi)+f \varphi'^2+m_0^2 \chi^2_{~2}
  \label{eq4:inseq:22},\\
  G^3_{~3} &=& G^4_4=\frac{1}{ 2 r R}[(r R) f'' + 2 (r R)' f'+ 2 (r R)''f]=-\frac{1}{2}V(\varphi)-f \varphi'^2+m_0^2 \chi^3_{~3}.
  \label{eq4:inseq:33}
\label{eq4:dilaton:1}
\end{eqnarray}
Here the prime $'$ denotes differentiation with respect to the radial coordinate $r$.

Based on Eqs.\eqref{eq4:chi:1}-\eqref{eq4:inseq:22},
 we obtain
 \begin{equation}\label{eq4:inseq:12}
   \frac{(r R)''}{r R}=-\varphi'^2,
 \end{equation}
which can be rewritten as the following form
\begin{equation}\label{eq4:lnR}
   \frac{d^2}{dr^2}\ln R +\frac{2}{r}\frac{d}{dr}\ln R +\left(\frac{d}{dr}\ln R\right)^2=-\varphi'^2.
 \end{equation}
In order to derive the dilaton field $\varphi$, we assume that $R(r)$ could be an exponential function of $\varphi(r)$, such as
\begin{eqnarray}\label{eq4:Rr}
  R(r)=e^{\alpha \varphi},
\end{eqnarray}
where $\alpha$ is a constant. Then, Eq.(\ref{eq4:lnR}) becomes a simple differential equation for $\varphi$
\begin{eqnarray}\label{eq4:dilaton:3}
\alpha \phi ''(r)+\left(\alpha ^2+1\right) \phi '(r)^2+\frac{2 \alpha  \phi '(r)}{r}=0.
\end{eqnarray}
In fact, the similar assumption (\ref{eq4:Rr}) has been extensively used to look for the charged dilaton black hole solutions~\cite{Dehghani:2004sa,Sheykhi:2007wg} in Maxwell-dilaton gravity.
By solving the Eq.(\ref{eq4:dilaton:3}), the dilaton field can be obtained as
\begin{equation}\label{eq4:varphi}
  \varphi (r) = \frac{\alpha}{1+\alpha^2} \ln\frac{\delta}{r}.
\end{equation}
Here the integration constant $\delta$ is supposedly related to some rescaling properties of solution.

Taking the trace of the gravitational field equation (\ref{eq4:eist}),
one can get
\begin{eqnarray}\label{eq4:contract}
f''&+&\frac{4( r R)'f'}{rR}+\frac{ 2 f\left((r R')^2+2 r R \left(r R''+3 R'\right)+R^2+(r R)^2\varphi '^2 \right)}{(r R)^2}\nonumber\\
&+&2 V(\varphi)-\frac{2}{(r R)^2}-m_0^2\sum^4_{i=1}\chi_i^i=0,
\end{eqnarray}
Considering the $G^3_3$ component of gravitational field equation (\ref{eq4:inseq:33}) together with the assumption (\ref{eq4:Rr}) and the solution of dilaton field (\ref{eq4:varphi}), Eq. (\ref{eq4:contract}) can be simplified as
\begin{eqnarray}
-\frac{f'}{\alpha ^2+1}
   +\frac{\left(\alpha   ^2-1\right) f}{\left(\alpha ^2+1\right)^2 r}
   -\frac{ r V(\varphi)}{2}
   +\frac{e^{\frac{1-\alpha^2}{\alpha}\varphi } }{\delta}
 + c_1 c_0 m_0^2 e^{-(\alpha+2 \beta_1)\varphi }
+\frac{ c_2 c_0^2 m_0^2 }{\delta} e^{\frac{1-\alpha^2-2\alpha \beta_2}{\alpha} \varphi}
=0
    \label{eq4:inseq:33:2}
\end{eqnarray}

On the other hand, we further consider the scalar field equation and substitute the metric ansatz \eqref{eq4:metric}  and scalar field \eqref{eq4:varphi} including Eqs.~\eqref{eq4:chi:1} \eqref{eq4:Rr} into Eq.~\eqref{eq4:dilaton}. Then, the scalar field equation becomes
\begin{eqnarray}
\frac{f'}{\alpha ^2+1}+\frac{\left(1-\alpha ^2\right) f}{(\alpha ^2 +1)^2 r}
  +\frac{ r}{4 \alpha }\frac{\partial V}{\partial \varphi}
  +\frac{  m_0^2 }{\alpha \delta}
  \left(\beta_2 c_2 c_0^2 e^{\frac{1-\alpha^2-2\alpha \beta_2}{\alpha} \varphi}
  + \beta_1 c_1 c_0\delta e^{-(\alpha+2 \beta_1) \varphi}\right)=0
  \label{eq4:dilaton:2}
\end{eqnarray}

According to Eqs. (\ref{eq4:inseq:33:2})
and (\ref{eq4:dilaton:2}), we obtain a first order differential equation for dilaton field potential
\begin{eqnarray}
\frac{ \partial V(\varphi )}{\partial \varphi }-2 \alpha  V(\varphi  )
+\frac{4 \alpha  e^{\frac{2 \varphi  }{\alpha }}}{\delta ^2}
+\frac{4  m_0^2 }{\delta^2 }
\left(c_1  \delta(\alpha +\beta_1) e^{-2 \varphi \left(\beta_1-\frac{1}{2 \alpha }\right)}
+ c_2 (\alpha +\beta_2) e^{-2 \varphi   \left(\beta_2-\frac{1}{\alpha }\right)}\right)=0
\label{eq4:potential :dilation}
\end{eqnarray}
\textbf{The solution to the differential  equation(\ref{eq4:potential :dilation}) can be written as the
generalized form of the Liouville scalar potential}
\begin{eqnarray}\label{eq4:di potential}
  V(\varphi)=\left\{
  \begin{array}{c}
  2\gamma_0 e^{2\xi_0 \varphi}
  +2 \gamma_{1} e^{2\xi_{1}\varphi}
  +2 \gamma_{2}e^{2\xi_{2}\varphi},
  \ \alpha\neq 1\\
   2\lambda_0\varphi e^{2 \varphi}
  +2 \lambda_{1} e^{2\zeta_{1}\varphi}
  +2 \lambda_{2}e^{2\zeta_{2}\varphi},
  \ \alpha= 1
   \end{array}
    \right.
\end{eqnarray}
where the last two terms are associating with coupling between the dilaton field and graviton,
and the parameters are determined as following
\begin{eqnarray}
  \gamma_0&=&\frac{ \alpha ^2 }{\left(\alpha ^2-1\right) \delta ^2},\ \
   \gamma_1=\frac{2 \alpha c_1 m_0^2 \left(\alpha +\beta _1\right) }{\delta  \left(2 \alpha ^2+2 \alpha  \beta _1-1\right)},\ \
\gamma_2=\frac{ \alpha c_2 m_0^2 \left(\alpha +\beta _2\right)}{\delta ^2 \left(\alpha ^2+\alpha  \beta _2-1\right)}
 \nonumber\\
 \xi_0&=&\frac{1}{\alpha}, \ \
  \xi_{1}=-\beta_{1}+ \frac {1}{2\alpha}, \ \
  \xi_{2}=-\beta_{2}+ \frac {1}{\alpha}, \zeta_1=1/2 -\beta _1,\ \ \zeta_2=1- \beta _2\nonumber\\
 \lambda_0&=&-\frac{2 }{\delta ^2},\ \ \lambda_1=\frac{2 \left(\beta _1+1\right) c_1 m_0^2}{2 \beta _1 \delta +\delta },\ \ \lambda_2=\frac{ \left(\beta _2+1\right) c_2 m_0^2}{\beta _2 \delta ^2}.
\end{eqnarray}
Then, the black hole solution $f(r)$ from Eq.(\ref{eq4:dilaton:2}) can be written as
\begin{eqnarray}\label{eq4:metr fina}
  f(r)=\left\{
  \begin{array}{l}
  -m r^{\frac{\alpha^2-1}{\alpha^2+1}}
   -\frac{\left(\alpha^2+1\right)  \delta^{-\frac{2   \alpha^2}{\alpha^2+1}} r^{\frac{2 \alpha^2}{\alpha^2+1}}}{\alpha^2-1}
   -\frac{\left(\alpha^2+1\right)^2 c_1 m_0^2
   \delta ^{-\frac{\alpha(\alpha +2\beta_1)}{\alpha ^2+1}}
   r^{\frac{2 \alpha^2+2\beta_1 \alpha +1}{\alpha ^2+1}}}
   {\left(\alpha ^2+2\beta_1 \alpha +2\right) \left(2 \alpha ^2+2\beta_1 \alpha   -1\right)}\\
  \ \ \ \ -\frac{ \left(\alpha ^2+1\right)^2 c_2  m_0^2
   \delta ^{\frac{-2\alpha(\beta_2+ \alpha )}{\alpha ^2+1}}
   r^{\frac{2\alpha  ( \alpha +\beta_2)}{\alpha^2+1}}}
   {\left(\alpha ^2+2\beta_2 \alpha +1\right) \left(
   \alpha ^2+\beta_2 \alpha -1\right)},  \qquad \alpha  \neq 1\\
  -m+\frac{2 r }{\delta }\left(\log \left(\frac{\delta }{r}\right)+2\right)-\frac{4 c_1 m_0^2  \delta^{-\beta _1-\frac{1}{2}} r^{\beta _1+\frac{3}{2}}}{4 \beta _1^2+8 \beta _1+3}
  -\frac{2 c_2 m_0^2 \delta^{-\beta _2-1} r^{\beta _2+1}}{\beta _2^2  +\beta _2  }, \quad \alpha=1
   \end{array}
   \right.
\end{eqnarray}
where $m$ is an integration constant related to the mass of the black hole as it will be shown below.
Therefore, there exist two branch solutions for the dilatonic black holes in the massive Einstein-dilaton gravity. 

Notice that, in the absence of the dilaton field $(\alpha=0)$, the solution $f(r)$ in massive gravity reduces to
\begin{eqnarray}\label{eq:sol:no}
f(r)=1-\frac{m}{r}+\frac{1}{2} c_1m_0^2 r+c_2 m_0^2,
\end{eqnarray}
which was presented in Ref.\cite{Cai:2015tb}. Here we have set $c_0=1$.
Obviously, the solution $f(r)$ in Eq.\eqref{eq:sol:no} does not describe an asymptotically flat spacetimes unless $m_0=0$. 
For the dilatonic black hole solution $f(r)$ in the dilaton massive gravity, the dominant term of metric function $f(r)$ approaches
\begin{eqnarray} \label{eq:asymp}
\lim_{r\to \infty} f(r)=\left\{
  \begin{array}{l}
   -\frac{\left(\alpha^2+1\right)  \delta^{-\frac{2   \alpha^2}{\alpha^2+1}} r^{\frac{2 \alpha^2}{\alpha^2+1}}}{\alpha^2-1}
   -\frac{\left(\alpha^2+1\right)^2 c_1 m_0^2
   \delta ^{-\frac{\alpha(\alpha +2\beta_1)}{\alpha ^2+1}}
   r^{\frac{2 \alpha^2+2\beta_1 \alpha +1}{\alpha ^2+1}}}
   {\left(\alpha ^2+2\beta_1 \alpha +2\right) \left(2 \alpha ^2+2\beta_1 \alpha   -1\right)}\\
  \ \ \ \ -\frac{ \left(\alpha ^2+1\right)^2 c_2  m_0^2
   \delta ^{\frac{-2\alpha(\beta_2+ \alpha )}{\alpha ^2+1}}
   r^{\frac{2\alpha  ( \alpha +\beta_2)}{\alpha^2+1}}}
   {\left(\alpha ^2+2\beta_2 \alpha +1\right) \left(
   \alpha ^2+\beta_2 \alpha -1\right)},  \qquad \alpha  \neq 1\\
  \frac{2 r }{\delta }\left(\log \left(\frac{\delta }{r}\right)+2\right)-\frac{4 c_1 m_0^2  \delta^{-\beta _1-\frac{1}{2}} r^{\beta _1+\frac{3}{2}}}{4 \beta _1^2+8 \beta _1+3}
  -\frac{2 c_2 m_0^2 \delta^{-\beta _2-1} r^{\beta _2+1}}{\beta _2^2  +\beta _2  }, \quad \alpha=1
   \end{array}
   \right.
\end{eqnarray}
at the infinity. For example,
taking $\alpha =0.6$, $\beta_1=0.5$ and $\beta_2 =0.3$, we have
\begin{equation}
    \lim_{r\to \infty} f(r)=2.125 r^{0.529412}-1.9527 c_1 m_0^2 r^{1.70588}+2.33771 c_2 m_0^2 r^{0.794118},
\end{equation}
where we set $\delta=1$ for simplify. Clearly, the metric function is also not asymptotically flat but asymptotic infinity in general.

For spacetime singularities, we calculate the Ricci and Kretschmann scalars
\begin{eqnarray}
\mathcal {R}\ \  &=&-\frac{4 f' (r R)'}{r R}-\frac{f \left(2 \left((r R)'\right)^2+4 r R (r R)''\right)}{(r R)^2}+\frac{2}{(r R)^2}-f''(r),\\
  \mathcal {R}^{\mu\nu\rho\sigma} \mathcal {R}_{\mu\nu\rho\sigma}
   &=&f \left(\frac{8 f' (r R)' (r R)''}{(r R)^2}-\frac{8 \left((r R)'\right)^2}{(r R)^4}\right)+\frac{4 f'^2 \left((r R)'\right)^2}{(r
   R)^2}\nonumber\\
   & &+\frac{4 f^2 \left(\left((r R)'\right)^4+2 (r R)^2 \left((r R)''\right)^2\right)}{(r R)^4}+\frac{4}{(r R)^4}+f''^2.
\end{eqnarray}
Note that both of Ricci and Kretschmann scalars are not singular at the horizons, therefore these points are just singularity of coordinate as it should be for a black hole. Considering the leading terms of asymptotical behaviors of metric at the origin, we obtain
\begin{eqnarray}
  \lim_{r \rightarrow 0} \mathcal {R} &\sim&  m r^{-\frac{\alpha ^2+3}{\alpha ^2+1}},  \ \ \ \ \ \ \ \lim_{r\rightarrow0} \mathcal {R}^{\mu\nu\rho\sigma}\mathcal {R}_{\mu\nu\rho\sigma}\sim  m r^{-\frac{\alpha ^2+5}{\alpha ^2+1}},
\end{eqnarray}
where Ricci and Kretschmann scalars are divergence at
origin $r=0$, finite for $r>0$, which suggests the origin $r=0$  is an essential and physical singularity in the spacetime.

In order to comprehend the behaviour of the metric function deeply, we would like to give graphical dependence of the function $f(r)$ both for $\alpha\neq 1$ and $\alpha =1$, and we set  $m_0=\delta=m=1$ for simplify in following discussions.

\subsection{$\alpha\neq 1$}

\begin{figure}[H]
  \subfigure[]{\label{fig:1:1} 
  \includegraphics[width=6cm]{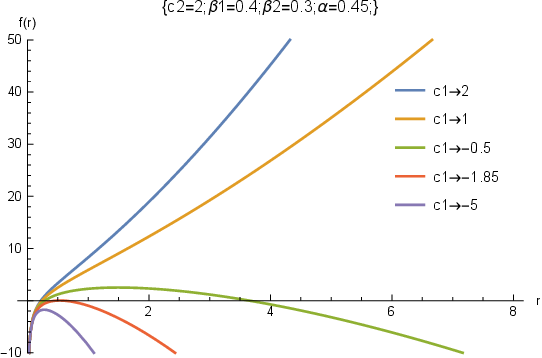}}
  \hfill%
  \subfigure[]{\label{fig:2:1} 
  \includegraphics[width=6cm]{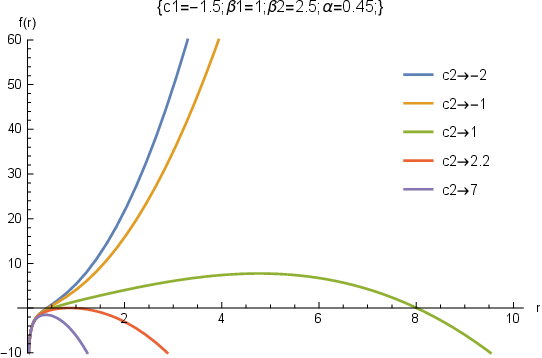}}
  \caption{The function $f(r)$ versus $r$ for different values of $c_1$  and $c_2$ in case $\alpha \neq 1$ }
    \label{fig:1}
\end{figure}

\begin{figure}[H]
  \subfigure[]{\label{fig:3:1} 
  \includegraphics[width=5cm]{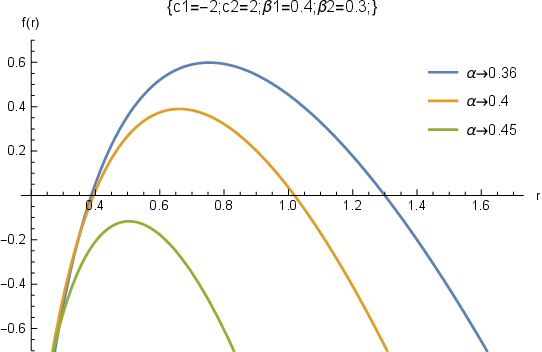}}
  \subfigure[]{\label{fig:3:2} 
  \includegraphics[width=5cm]{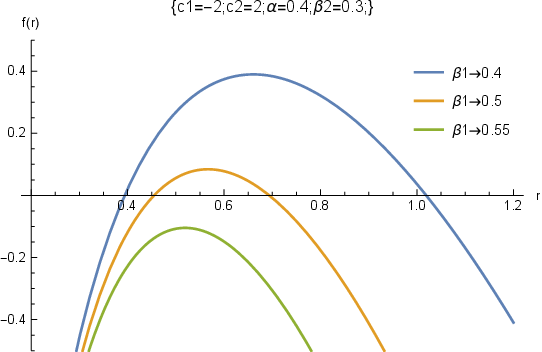}}
  \subfigure[]{\label{fig:3:3} 
  \includegraphics[width=5cm]{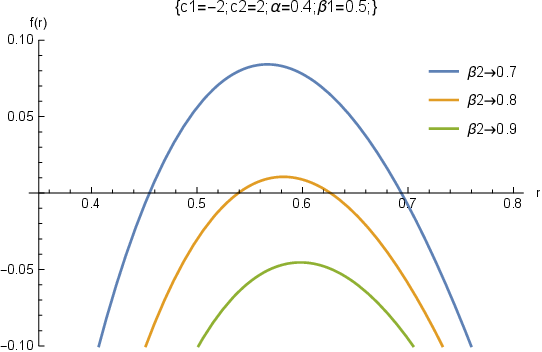}}
  \caption{The function $f(r)$ versus $r$ for different values of $\alpha$, $\beta_1$ and $\beta_2$ in the case $\alpha \neq 1$ }
    \label{fig:2}
\end{figure}

\textbf{For $\alpha \neq 1$, Figs. \ref{fig:1} and \ref{fig:2} depict the plots of $f(r)$ versus $r$, indicating the presence of black hole horizons.
Fig.\ref{fig:1:1} shows that when $c_1>0$ and $c_2>0$, there is only one black hole horizon, and $f(r)$ approaches to $+\infty$ if $r\rightarrow\infty$.  However, when $c_1<0$, there are two horizons: an event horizon and a cosmological horizon, and $f(r)$ approaches to $-\infty$ as $r\rightarrow\infty$.
As $c_1$ decreases to a certain value, 
an extremal black hole known as the Nariai black hole is formed, exhibiting a coincidence of the event and cosmological horizons. A naked singularity may also appear with the decreasing of $c_1$. Similarly, Fig.\ref{fig:2:1} shows that when $c_2<0$ and $c_1<0$, there is only one black hole horizon, but when $c_2>0$, there are two horizons: an event horizon and a cosmological horizon. 
It's worthy noting that as shown in Eq. (\ref{eq:sol:no}) for black hole solution in massive gravity, the parameter $c_2$ does not affect the asymptotic behavior of the solution $f(r)$. However, with the nonminimal coupling to the dilaton field, $c_2$ plays an important role in the behavior of solution $f(r)$ because the term of  $c_2$ can become the dominant term with suitable parameters $\alpha$, $\beta_1$, and $\beta_2$ from Eq. \eqref{eq:asymp}.
The influences of the parameters $\alpha$, $\beta_1$, and $\beta_2$ are also respectively depicted in Fig.\ref{fig:2}.}

\subsection{$\alpha= 1$}

\begin{figure}[H]
  \subfigure[]{\label{fig:4:1} 
  \includegraphics[width=6cm]{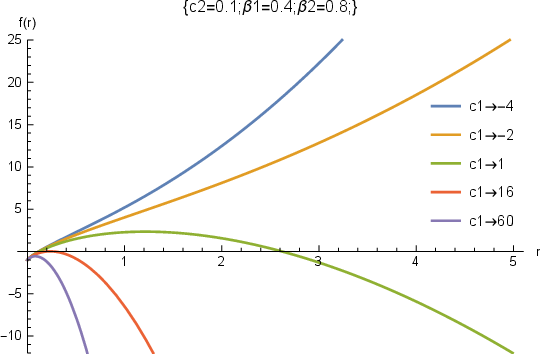}}
  \hfill%
  \subfigure[]{\label{fig:5:1} 
  \includegraphics[width=6cm]{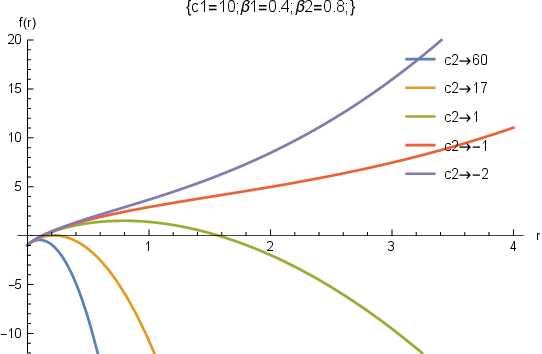}}
  \caption{The function $f(r)$ versus $r$ for different values of $c_1$  and $c_2$ in case $\alpha=1$ }
    \label{fig:3}
\end{figure}

\begin{figure}[H]
  \subfigure[]{\label{fig:6:1} 
  \includegraphics[width=6cm]{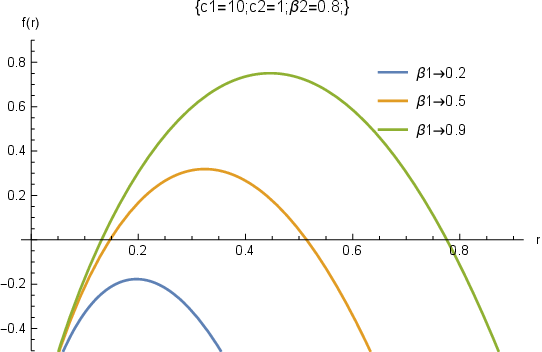}}
  \hfill%
  \subfigure[]{\label{fig:6:2} 
  \includegraphics[width=6cm]{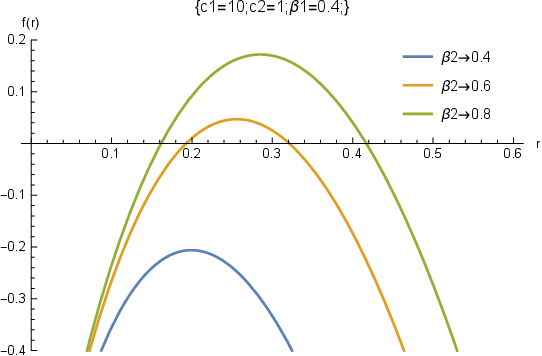}}
  \caption{The function $f(r)$ versus $r$ for different values of $\beta_1$ and $\beta_2$ in case $\alpha =1$ }
    \label{fig:4}
\end{figure}

When $\alpha = 1$, there is still the presence of one black hole horizon with or without a cosmological horizon
for different asymptotical behaviors, 
as shown in Fig. \ref{fig:3}.
The influences of other parameters  $\beta_1$ and $\beta_2$, 
are separately demonstrated in Figs.\ref{fig:4}. 
Similarly to the case when $\alpha \neq 1$, 
it is possible to have an extremal black hole, 
known as the Nariai black hole, by selecting appropriate parameter values. 
Additionally, changing one of the parameters can lead to the appearance of a naked singularity.

\section{THERMODYNAMICS OF DILATONIC BLACK HOLES}
\label{3s}

\begin{figure}[H]
  \subfigure[]{\label{fig:m1:1} 
  \includegraphics[width=6cm]{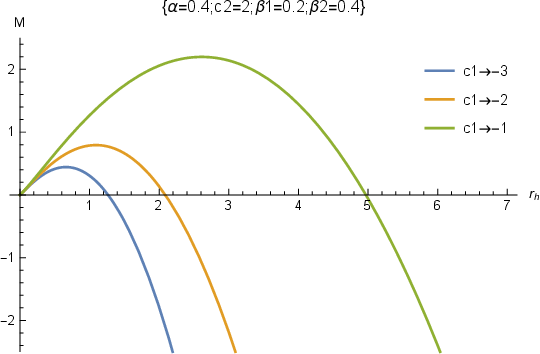}}
  \subfigure[]{\label{fig:m1:2} 
  \includegraphics[width=6cm]{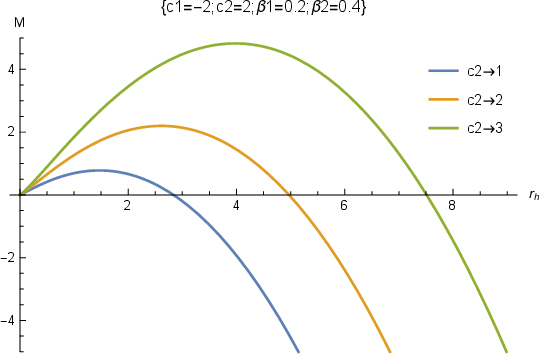}}
  \subfigure[]{\label{fig:m1:3} 
  \includegraphics[width=6cm]{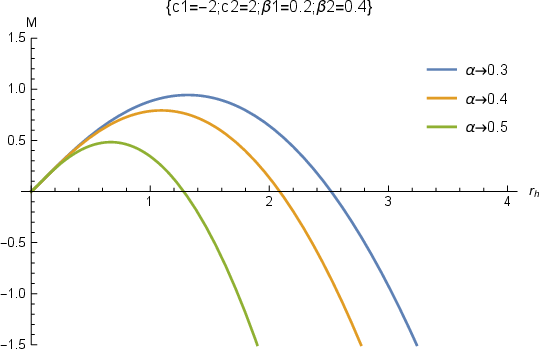}}\\
  \subfigure[]{\label{fig:m1:4} 
  \includegraphics[width=6cm]{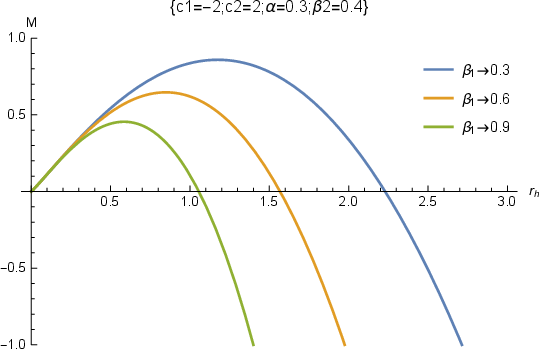}}
  \subfigure[]{\label{fig:m1:5} 
  \includegraphics[width=6cm]{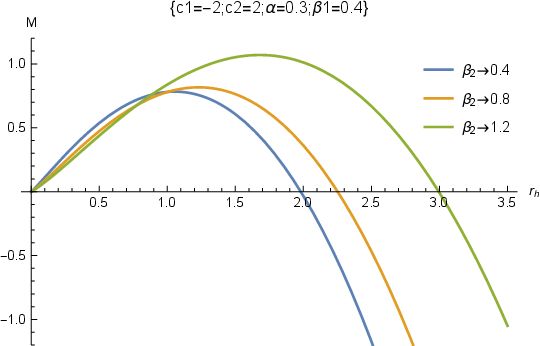}}
  \caption{The mass $M$ of black hole versus $r_h$  in the case $\alpha \neq 1$}
    \label{fig:m1}
\end{figure}

\begin{figure}[H]
  \subfigure[]{\label{fig:m3:1} 
  \includegraphics[width=6cm]{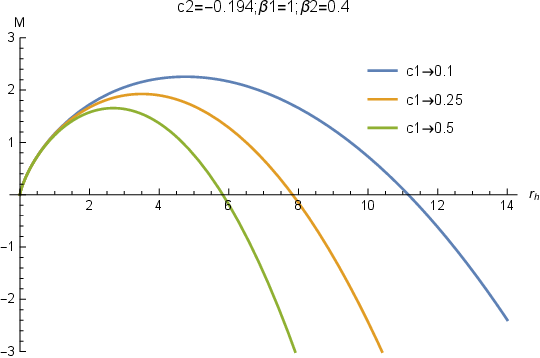}}
  \hfill%
  \subfigure[]{\label{fig:m3:2} 
  \includegraphics[width=6cm]{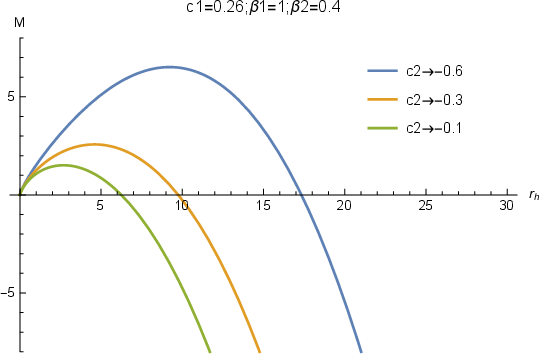}}
  \hfill
  \subfigure[]{\label{fig:m3:3} 
  \includegraphics[width=6cm]{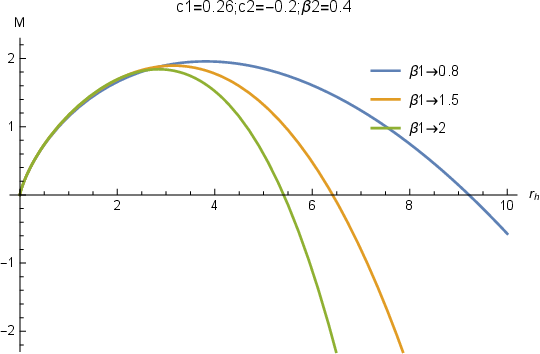}}
  \hfill%
  \subfigure[]{\label{fig:m3:4} 
  \includegraphics[width=6cm]{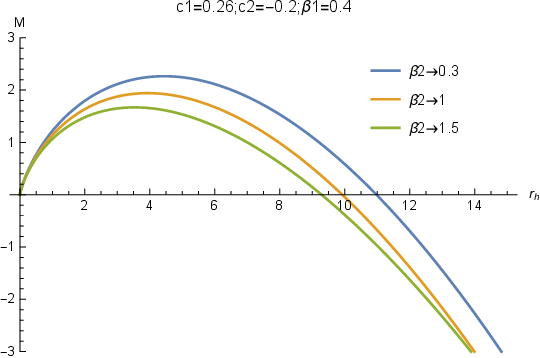}}
  \caption{The mass $M$ of black hole versus $r_h$  in the case $\alpha =1$}
    \label{fig:m3}
\end{figure}

Now we plan to investigate the thermodynamics of theses dilatonic black holes in massive Einstein-dilaton gravity.
According to the definition of ADM mass~\cite{Abbott:1982} and definition of horizon~ $f(r_h)=0$, the mass of dilatonic black hole is given by
\begin{eqnarray}\label{eq4:mass}
  M&=&\frac{\delta^{\frac{2\alpha^2}{1+\alpha^2}}m}{2(1+\alpha^2)}\nonumber\\
   &=& \left\{\begin{array}{l}
                \frac{r_h}{2(1- \alpha ^2)}
   -\frac{\left(\alpha ^2+1\right) c_0 c_1 m_0^2 \delta ^{\frac{\alpha  (\alpha -2 \beta_1)}{\alpha ^2+1}} r_h^{\frac{\alpha ^2+2 \alpha  \beta_1+2}{\alpha ^2+1}}}{2 \left(\alpha ^2+2 \alpha  \beta_1+2\right) \left(2 \alpha ^2+2 \alpha
   \beta_1-1\right)}
   -\frac{\left(\alpha ^2+1\right) c_0^2 c_2 m_0^2 \delta ^{-\frac{2 \alpha  \beta_2}{\alpha ^2+1}} r_h^{\frac{\alpha ^2+2 \alpha  \beta_2+1}{\alpha ^2+1}}}{2 \left(\alpha ^2+\alpha  \beta_2-1\right) \left(\alpha ^2+2 \alpha
   \beta_2+1\right)},\ \alpha\neq 1 \\
   \frac{1}{2} r_h \left(\log \left(\frac{\delta }{r_h}\right)+2\right)
   -\frac{c_0 c_1 m_0^2 r_h^2 \left(\frac{\delta }{r_h}\right)^{\frac{1}{2}-\beta_1}}{4 \beta_1 (\beta_1+2)+3}
   -\frac{c_0^2 c_2 m_0^2 r_h \left(\frac{\delta }{r_h}\right)^{-\beta_2}}{2 \left(\beta_2^2+\beta_2\right)}
                ,\ \alpha=1
              \end{array}
\right.
\end{eqnarray}
The ADM mass as a function of black hole radius are plotted in Figs. \ref{fig:m1} and \ref{fig:m3} for both cases of $\alpha\neq1$, and $\alpha=1$. In each case,  there exists a maximum mass $M_{max}$,  and there are two black holes with the same mass,  distinguished by their size (a smaller one and a larger one).  The ADM mass is always positive, 
and therefore there is also a maximum radius $r_{max}$ at which the mass vanishes.
It can be observed that for $\alpha\neq1$, 
the maximum  mass $M_{max}$ increases with $c_1$ and $c_2$ 
but decreases with $\alpha$ (see Figs. \ref{fig:m1:1}- \ref{fig:m1:3}). 
Conversely, $M_{max}$ decreases with $c_1$ and $c_2$ when $\alpha\neq 1$
(see Figs.\ref{fig:m3:1} and \ref{fig:m3:2}).
The remaining figures demonstrate the effects of the the coupling of dilaton field. 
The maximum  mass always decreases with $\beta_1$ for both cases of $\alpha$, 
as shown in Figs.\ref{fig:m1:4} and \ref{fig:m3:3}, 
suggesting that the coupling of dilaton field to the first terms of the graviton  
could always reduced the  maximum  mass of black holes.
However, for $\beta_2$, $M_{max}$  increases when $\alpha \neq 1$ but decreases when $\alpha=1$
(see Figs. \ref{fig:m1:5} and \ref{fig:m3:4}).

\begin{figure}[H]
  \subfigure[]{\label{fig:t1:1} 
  \includegraphics[width=6cm]{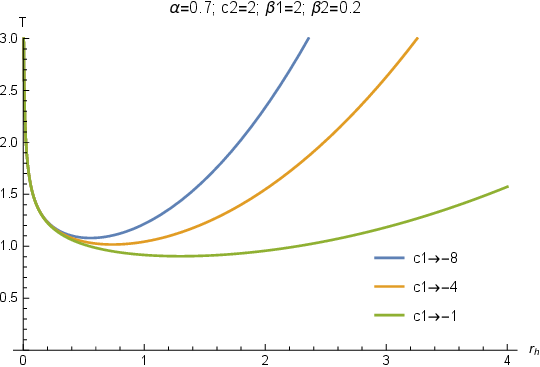}}
  \subfigure[]{\label{fig:t1:2} 
  \includegraphics[width=6cm]{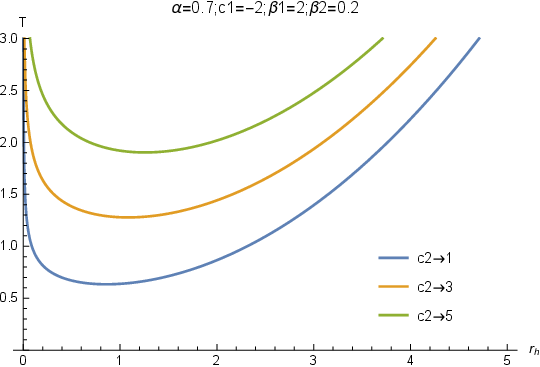}}
  \subfigure[]{\label{fig:t1:3} 
  \includegraphics[width=6cm]{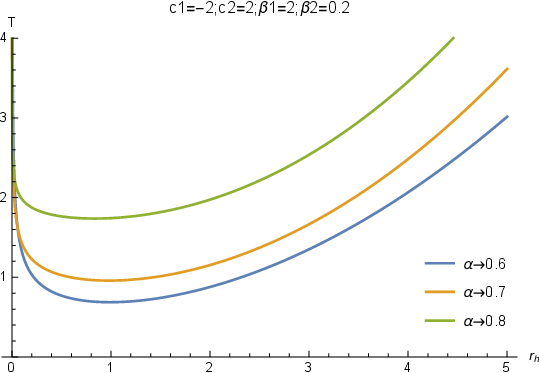}}\\
  \subfigure[]{\label{fig:t1:4} 
  \includegraphics[width=6cm]{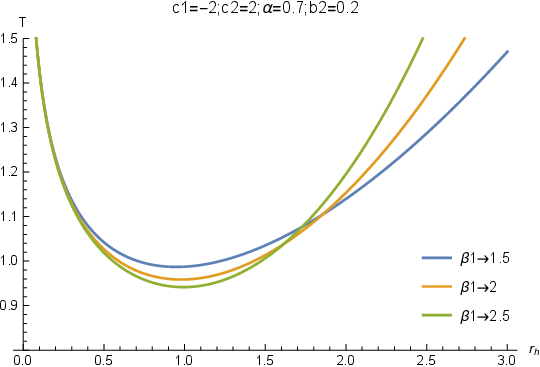}}
  \subfigure[]{\label{fig:t1:5} 
  \includegraphics[width=6cm]{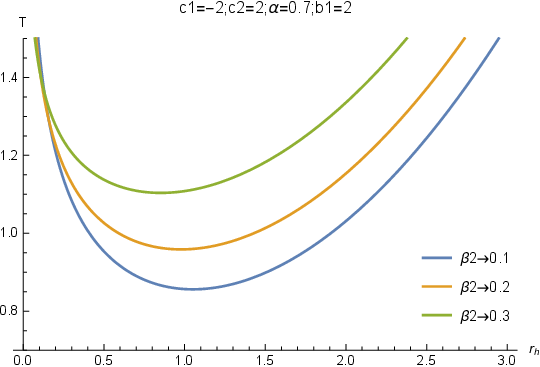}}
  \caption{The temperature $T$ versus $r_h$ for $\alpha\neq1$}
    The extreme points of each $T-r$ curves ( $r_{cri}$, $T_{min}$ ) are,  respectively,
  (a) (0.555418, 1.07929), (0.737029, 1.01689), (1.29893, 0.90413) for $c_1 = -8, -4, -1$
  (b) (0.858517, 0.633958), (1.08183, 1.27758), (1.25738, 1.90305) for $c_2 = 1, 3, 5$
  (c) (0.989441, 0.687425), (0.978303, 0.958608), (0.833922, 1.73715) for $\alpha = 0.6, 0.7, 0.8$
  (d) (0.948498, 0.986909), (0.978303, 0.958608), (0.992489, 0.941541) for $\beta_1 = 1.5, 2, 2.5$
  (e) (1.05142, 0.856112), (0.978303, 0.958608), (0.852074, 1.10364)  for $\beta_2 = 0.1, 0.2, 0.3$
  \label{fig:t1}
\end{figure}

\begin{figure}[H]
  \subfigure[]{\label{fig:t3:1} 
  \includegraphics[width=4cm]{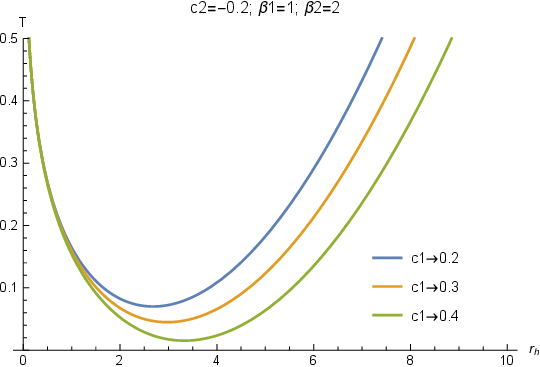}}
  \subfigure[]{\label{fig:t3:2} 
  \includegraphics[width=4cm]{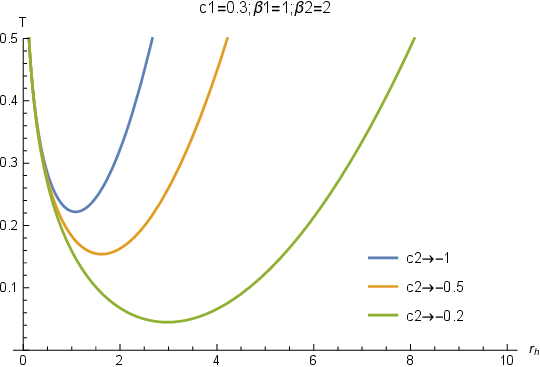}}
  \subfigure[]{\label{fig:t3:3} 
  \includegraphics[width=4cm]{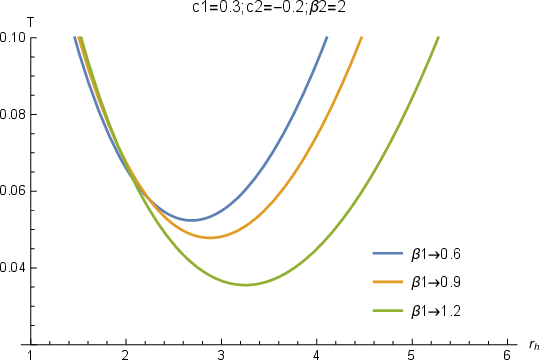}}
  \subfigure[]{\label{fig:t3:4} 
  \includegraphics[width=4cm]{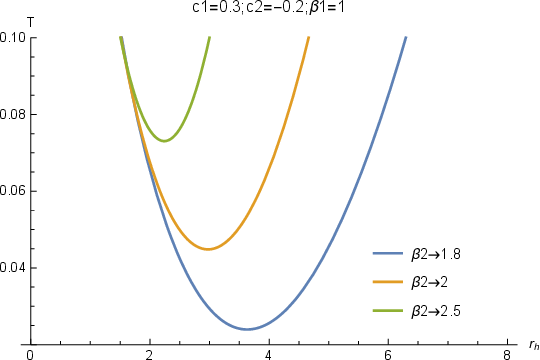}}
  \caption{The temperature $T$ versus $r_h$ for $\alpha=1$}
  The extreme points of each $T-r$ curves ( $r_{cri}$, $T_{min}$ ) are,  respectively,
  (a) (2.68273, 0.0700174), (2.97447, 0.0448315), (3.32688, 0.0152299) for $c_1 = 0.2, 0.3, 0.4$
  (b) (1.08101, 0.221862), (1.61782, 0.153979), (2.97447, 0.0448315) for $c_2 = -1, -0.5, -0.2$
  (c) (2.68804, 0.0523771), (2.88079, 0.047834), (3.24997, 0.0355224) for $\beta_1 = 0.6, 0.9, 1.2$
  (d) (3.63175, 0.0239511), (2.97447, 0.0448315), (2.24071, 0.0730591)  for $\beta_2 = 1.8, 2, 2.5$
  \label{fig:t3}
\end{figure}

To develop thermodynamics of dilatonic black hole, we need to calculate the Hawking temperature of the black hole geometrically associated with the black hole horizon. In terms of the surface gravity $\kappa$ corresponding to  the null killing
vector $(\frac{\partial}{\partial t})^a$  at the horizon, the temperature can be written as
\begin{eqnarray}
T &=& \frac{\kappa}{2 \pi}=\frac{1}{4 \pi}\frac{\partial f(r)}{\partial r}\big|_{r=r_h}\\ \nonumber
  &=&\left\{
  \begin{array}{l}
  -\frac{\alpha ^2+1}{4 \pi  r_h}\left(\frac{ \delta ^{-\frac{2 \alpha ^2}{\alpha ^2+1}} r_h^{\frac{2 \alpha ^2}{\alpha   ^2+1}}}{\alpha ^2-1}+\frac{c_0 c_1 m_0^2 \delta ^{-\frac{\alpha  \left(\alpha +2 \beta _1\right)}{\alpha ^2+1}} r_h^{\frac{2 \alpha ^2+2 \alpha  \beta _1+1}{\alpha ^2+1}}}{2 \alpha  \left(\alpha +\beta _1\right)-1}+\frac{c_0^2 c_2 m_0^2   \delta ^{-\frac{2 \alpha  \left(\alpha +\beta _2\right)}{\alpha ^2+1}} r_h^{\frac{2 \alpha  \left(\alpha +\beta _2\right)}{\alpha ^2+1}}}{\alpha  \left(\alpha +\beta _2\right)-1}\right),
  \ \alpha\neq 1 \\
  \frac{\log \left(\frac{\delta }{r_h}\right)+1}{2 \pi  \delta }
  -\frac{c_0 c_1 m_0^2 r_h \left(\frac{\delta }{r_h}\right){}^{\frac{1}{2}-\beta_1}}{2 \pi  (2 \beta_1+1) \delta }
  -\frac{c_0^2 c_2 m_0^2 \left(\frac{\delta }{r_h}\right){}^{-\beta_2}}{2 \pi  \beta_2 \delta } ,\ \alpha=1
            \end{array}
\right.
   \label{eq4:temp}
\end{eqnarray}
The Temperature of black hole as a function of the radius is plotted in Figs. \ref{fig:t1} and \ref{fig:t3} for different cases: $\alpha\neq1$ and $\alpha=1$.
For both  cases, there is a minimum positive temperature $T_{min}$ of the black hole, 
which illustrates there exist two black holes with the same temperature, distinguished by their size (a smaller one and a larger one).
Moreover, the minimum temperature ($T_{min}$ ) decreases with $c_1$ and $\beta_1$ (see Figs. \ref{fig:t1:1}, \ref{fig:t1:4}, \ref{fig:t3:1}, and \ref{fig:t3:3}). This suggests that the first term of the graviton,  including the effect of coupling to the dilaton field, always reduces the minimum  temperature of the black holes.
Figure.\ref{fig:t3:2} demonstrates that the minimum temperature increases with $c_2$ when $\alpha \neq 1$ but decreases with $c_2$ when $\alpha =1$.
For $\alpha \neq 1$ (as shown in Figs. \ref{fig:t1:3}, the minimum temperature increases with $\alpha$, indicating that the dilaton field could always enlarge the minimum  temperature.
In addition, also for both cases, the minimum temperature decreases with $\beta_1$ but increases with $\beta_2$, illustrating  that the coupling of dilaton field to the first term of graviton could always reduce the minimum temperature of black hole but the second term of graviton could always enlarge that of black hole (as shown in Figs. \ref{fig:t1:4},  \ref{fig:t1:5}, \ref{fig:t3:3}, and \ref{fig:t3:4}).

The entropy of dilatonic black hole is given by
\begin{eqnarray}  \label{eq4:entropy}
S=\pi r_h^2R(r_h)^2=\pi  \delta ^{\frac{2 \alpha ^2}{\alpha ^2+1}} r_h^{\frac{2}{\alpha ^2+1}}.
\end{eqnarray}
Refs.~\cite{Babichev:2014ac}-\cite{Zou:2016sab} have pointed out that the graviton mass does not significantly affect the form of
the entropy, and contributes only as a correction for the horizon radius. In fact, the entropy of these black holes in massive Einstein-dilaton gravity can be also derived from traditional metric ansatz of dilatonic black holes, as shown in Appendix. Then, we find that the thermodynamic quantities satisfy the first law of black hole thermodynamics
\begin{eqnarray}\label{eq4: first law}
  d M = T dS.
\end{eqnarray}

In addition, another important aspect of black hole thermodynamics involves analyzing the thermal stability of black hole solutions. To assess the thermal stability, we compute the heat capacity
\begin{eqnarray}
  C=T \frac{\partial S}{\partial T} = T \frac{\partial S/\partial r_h}{\partial T/\partial r_h},
  \label{eq4:capcty}
\end{eqnarray}
which leads to
\begin{eqnarray}
C &=&\left\{
\begin{array}{l}\frac{-8 \pi ^2 (\alpha ^2+1)^{-1} T \delta ^{\frac{4 \alpha ^2}{\alpha ^2+1}} r^{-\frac{2 \alpha ^2}{\alpha ^2+1}}
   r_h^{\frac{2}{\alpha ^2+1}+1}}{1+\frac{\alpha  c_0 c_1 m_0^2 \left(\alpha +2 \beta _1\right) \delta ^{\frac{\alpha  \left(\alpha -2 \beta
   _1\right)}{\alpha ^2+1}} r^{\frac{2 \alpha  \beta _1+1}{\alpha ^2+1}}}{2 \alpha ^2+2 \alpha  \beta
   _1-1}+\frac{c_0^2 c_2 m_0^2 \left(\alpha ^2+2 \alpha  \beta _2-1\right) \delta ^{-\frac{2 \alpha  \beta
   _2}{\alpha ^2+1}} r^{\frac{2 \alpha  \beta _2}{\alpha ^2+1}}}{\alpha ^2+\alpha  \beta _2-1}},\alpha\neq 1 \\
  -\frac{4 \pi ^2 T \delta ^{\beta _1+\beta _2+\frac{5}{2}} r_h^{3/2}}{c_0 c_1 m_0^2 \delta ^{\beta _2+1} r_h^{\beta _1+1}+2 c_0^2 c_2 m_0^2 \delta ^{\beta _1+\frac{1}{2}} r_h^{\beta _2+\frac{1}{2}}+2 \delta ^{\beta _1+\beta _2+\frac{1}{2}} \sqrt{r_h}},\ \alpha=1
\end{array}
\right.
  \label{eq4:cap}
\end{eqnarray}

\begin{figure}[H]
  \subfigure[]{\label{fig:cc1:1} 
  \includegraphics[width=5cm]{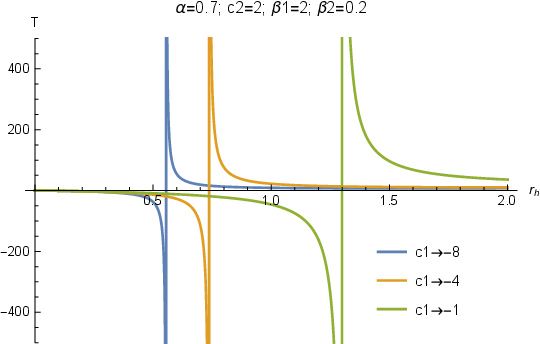}}
  \subfigure[]{\label{fig:cc1:2} 
  \includegraphics[width=5cm]{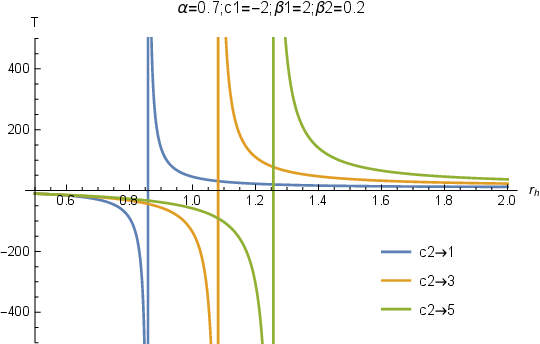}}
  \subfigure[]{\label{fig:cc1:3} 
  \includegraphics[width=5cm]{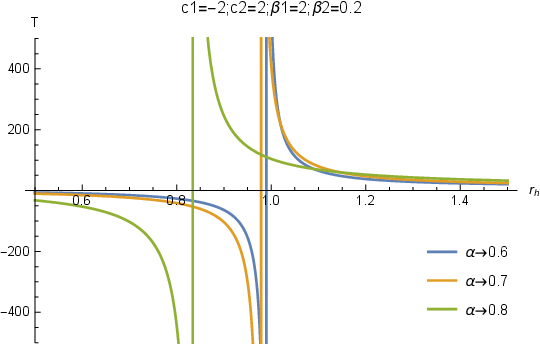}}\\
  \subfigure[]{\label{fig:cc1:4} 
  \includegraphics[width=5cm]{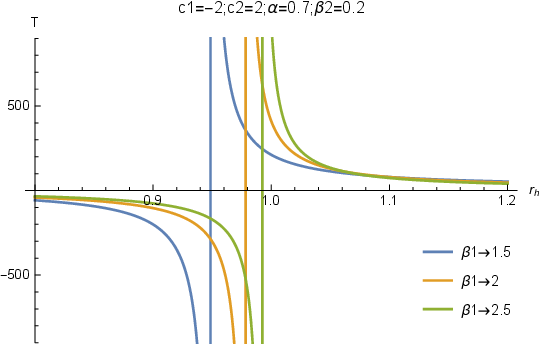}}
  \subfigure[]{\label{fig:cc1:5} 
  \includegraphics[width=5cm]{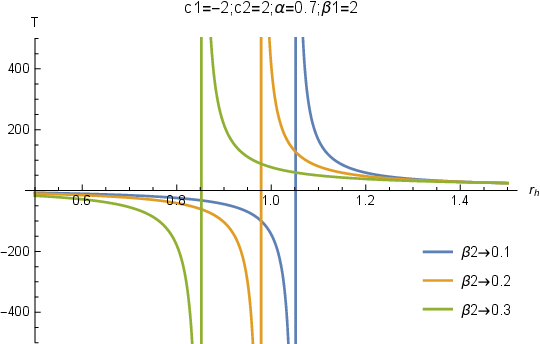}}
  \caption{The  heat capacity $C$ versus $r_h$ for $\alpha\neq1$}
  \label{fig:cc1}
\end{figure}

\begin{figure}[H]
  \subfigure[]{\label{fig:cc3:1} 
  \includegraphics[width=4cm]{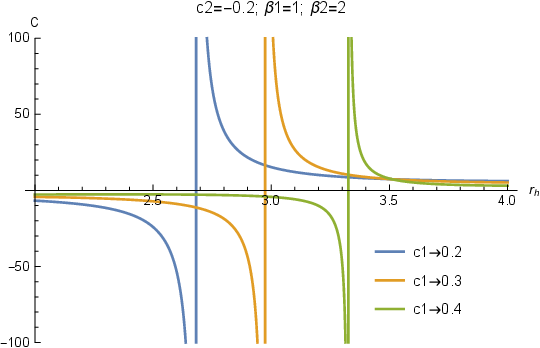}}
  \subfigure[]{\label{fig:cc3:2} 
  \includegraphics[width=4cm]{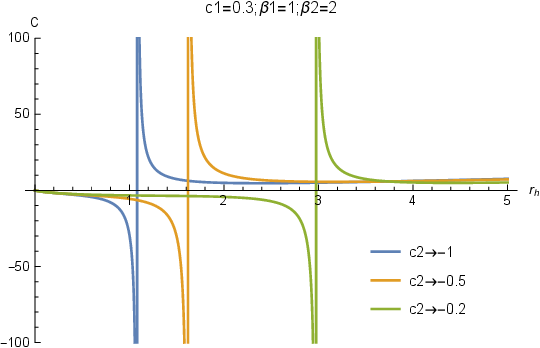}}
  \subfigure[]{\label{fig:cc3:3} 
  \includegraphics[width=4cm]{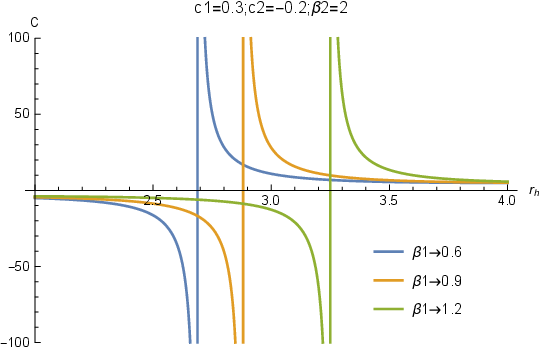}}
  \subfigure[]{\label{fig:cc3:4} 
  \includegraphics[width=4cm]{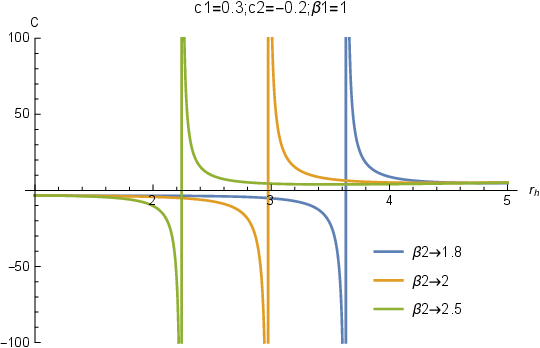}}
  \caption{The  heat capacity $C$ versus $r_h$ for $\alpha = 1$}
  \label{fig:cc3}
\end{figure}

From Eq.(\ref{eq4:capcty}),
the non-monotonic nature of the temperature allows us to
conclude that the heat capacity exhibits discontinuities  as illustrated in Figs.(\ref{fig:cc1} and \ref{fig:cc3}) for both cases.
Figs. \ref{fig:cc1} and \ref{fig:cc3} show that black holes with relatively large horizon radii ($r_h > r_{cri}$) are stable thermodynamic systems, while a domain of instability exists for smaller radii below the point of discontinuity ($r_h < r_{cri}$).
In other words, the figures clearly illustrate the stable and unstable regions based on the discontinuities observed in the heat capacity as a function of the horizon radius.

\begin{figure}[H]
  \subfigure[]{\label{fig:g1:1} 
  \includegraphics[width=5cm]{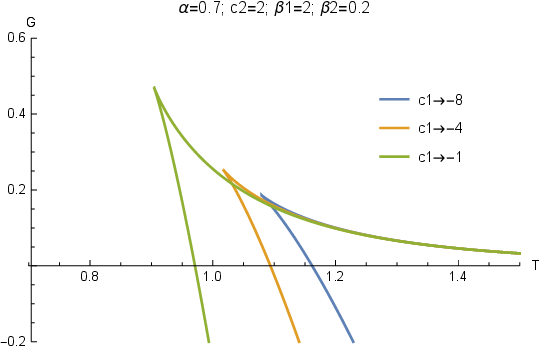}}
  \subfigure[]{\label{fig:g1:2} 
  \includegraphics[width=5cm]{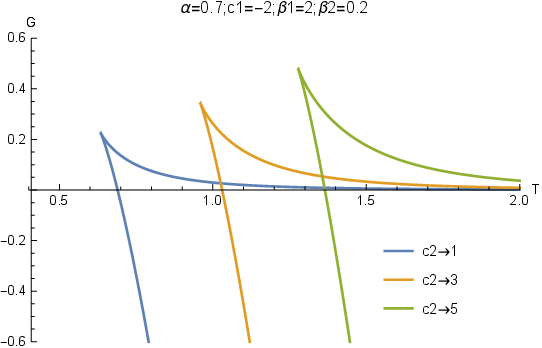}}
  \subfigure[]{\label{fig:g1:3} 
  \includegraphics[width=5cm]{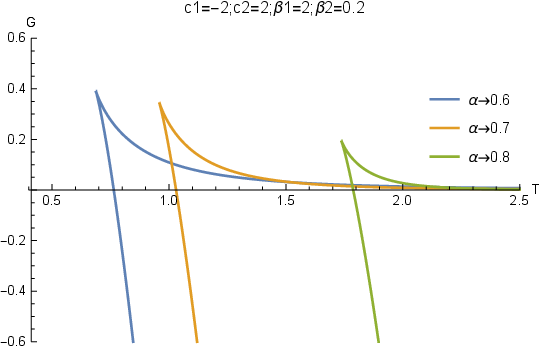}}\\
  \subfigure[]{\label{fig:g1:4} 
  \includegraphics[width=5cm]{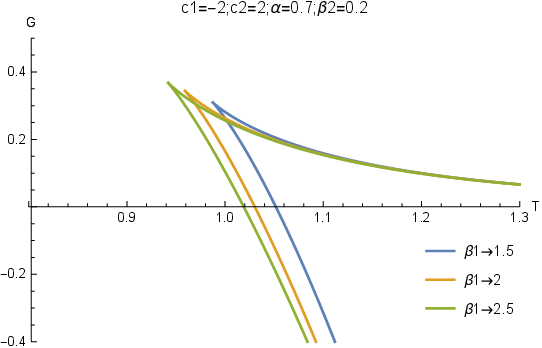}}
  \subfigure[]{\label{fig:g1:5} 
  \includegraphics[width=5cm]{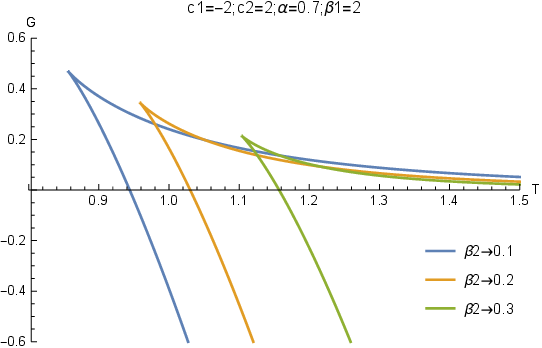}}
  \caption{The Gibbs free energy $G$ versus $r_h$ for $\alpha\neq1$}
    \label{fig:g1}
\end{figure}

\begin{figure}[H]
  \subfigure[]{\label{fig:g3:1} 
  \includegraphics[width=4cm]{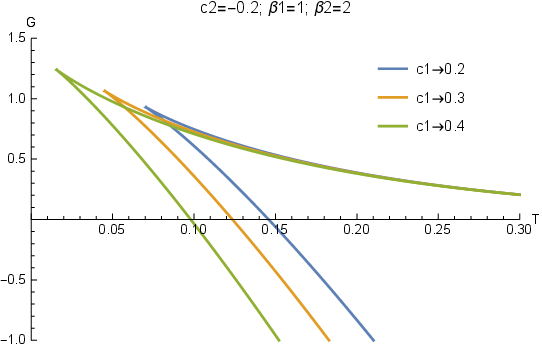}}
  \subfigure[]{\label{fig:g3:2} 
  \includegraphics[width=4cm]{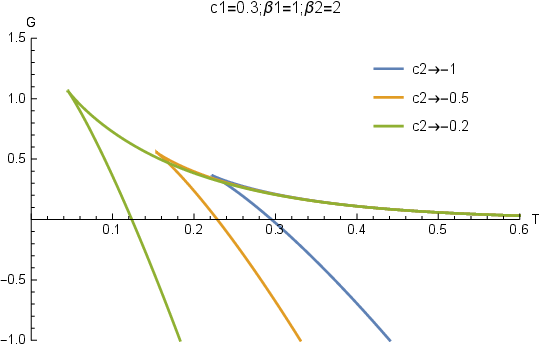}}
  \subfigure[]{\label{fig:g3:3} 
  \includegraphics[width=4cm]{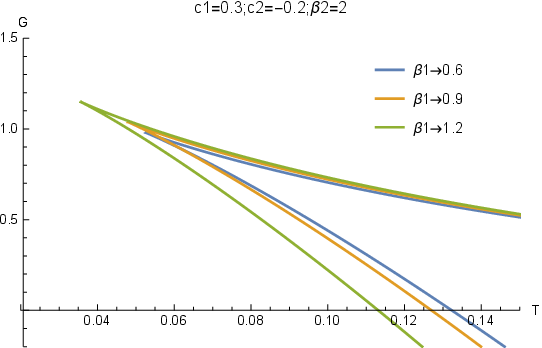}}
  \subfigure[]{\label{fig:g3:4} 
  \includegraphics[width=4cm]{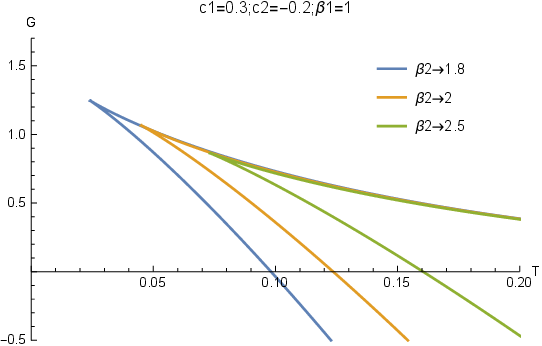}}
  \caption{The Gibbs free energy $G$ versus $r_h$ for $\alpha = 1$}
    \label{fig:g3}
\end{figure}

To get more information about the phase transition, we
can examine the free energy. The Gibbs free energy is $G=M-TS$ reads as
\begin{eqnarray}
G&=&\left\{
\begin{array}{l}
  \frac{r}{4}
+\frac{\alpha  \left(\alpha ^2+1\right) c_0  c_1 m_0^2 (\alpha +2 \beta_1) \delta ^{\frac{\alpha  (\alpha -2 \beta_1)}{\alpha ^2+1}} r^{\frac{\alpha ^2+2 \alpha  \beta_1+2}{\alpha ^2+1}}}{4 \left(\alpha ^2+2 \alpha \beta_1+2\right) \left(2 \alpha ^2+2 \alpha  \beta_1-1\right)}\\
+\frac{\left(\alpha ^2+1\right) c_0^2 c_2 m_0^2 \left(\alpha ^2+2 \alpha  \beta_2-1\right) \delta ^{-\frac{2 \alpha  \beta_2}{\alpha ^2+1}} r^{\frac{\alpha ^2+2 \alpha  \beta_2+1}{\alpha ^2+1}}}{4 \left(\alpha ^2+\alpha  \beta_2-1\right) \left(\alpha ^2+2 \alpha  \beta_2+1\right)},\
\alpha\neq 1 \\
  \frac{c_0 c_1 m_0^2 \delta ^{\frac{1}{2}-\beta _1} r_h^{\beta _1+\frac{3}{2}}}{4 \beta _1+6}+\frac{c_0^2 c_2 m_0^2 \delta ^{-\beta _2} r_h^{\beta _2+1}}{2 \beta _2+2}+\frac{r_h}{2},\ \alpha=1.
\end{array}
\right.
\label{eq4:gib}
\end{eqnarray}
We find that these discontinuities, which correspond to the minima of temperature in the heat capacity curves, 
indicate the occurrence of Hawking-Page phase transitions. 
Hawking-Page phase transitions were also recently found in dilatonic black holes 
that are neither asymptotically flat nor AdS as  in Ref.\cite{Dehghani:2019cuf}.
As shown in Figs.\ref{fig:g1} and \ref{fig:g3}, 
the upper and lower branches correspond to small and large black holes, respectively. 
A positive Gibbs free energy indicates that the system is in a radiation phase, 
whereas a Hawking-Page phase transition occurs at the intersection point of the lower branch with $G=0$. 
The temperature at this point is known as the Hawking-Page temperature, $T_{HP}$. 
The fact that large black holes always have a lower Gibbs free energy compared to small black holes 
confirms the above arguments regarding their thermal stability.
The effects of parameter variations can also be seen from Figs.\ref{fig:g1} and \ref{fig:g3}. 
Increasing $c_1$ or $\beta_1$ lowers $T_{HP}$. Similarly, lowering $\alpha$ or $\beta_2$ lowers the Hawking-Page temperature $T_{HP}$. 
However, $T_{HP}$ increases with $c_2$ when $\alpha\neq 1$ but decreases with $c_2$ when $\alpha=1$.

\section{BLACK HOLE SOLUTIONS IN $(d+1)$-DIMENSIONAL SPACETIME}
\label{4s}

In this section, we would like to extend to the massive Einstein-dilaton gravity in high dimensional spacetime.
The  action is given as
\begin{equation}\label{eqd:action}
  I=\frac{1}{16 \pi} \int d^{(d+1)} x \sqrt{-g}
  \Big[\mathcal {R}  -\frac{4}{d-1}(\nabla \varphi)^2-V(\varphi)
  + m_0^2 \sum_{i=1}^{4} c_i e^{-2\beta_i \varphi} \mathcal {U}_i(g,h)\Big],
\end{equation}
where the last term denotes  the general form of  the coupling between the scalar field and massive graviton, the  $\mathcal{U}_i$ are symmetric polynomials of the eigenvalues of the ~$(d+1)\times (d+1)$\ matrix
$K^{\mu}_{\nu}=\sqrt{g^{\mu\alpha}h_{\alpha\nu}}$, and satisfy the  relation in Eq.(\ref{eq4:u1234}).

By varying
the action (\ref{eqd:action}) with regard to metric $g_{\mu\nu}$ and $\varphi$, the equations of motions can be written as
\begin{eqnarray}
\mathcal {R}_{\mu\nu}-\frac{1}{2}\mathcal {R} g_{\mu\nu}
                    &=& \frac{4}{d-1} \partial_{\mu}\varphi\partial_{\nu}\varphi
                  -(\frac{V}{2}+\frac{2}{d-1} \partial^{\rho}\varphi\partial_{\rho}\varphi)g_{\mu\nu}
                  +m^2 \chi_{\mu\nu}, \label{eqd:eist} \\
  \nabla^2\varphi  &=& \frac{d-1}{8}\big[\frac{\partial V}{\partial \varphi}
                  -m^2 \sum_{i=1}^{4} \frac{\partial \tilde{c}_i}{\partial \varphi} \mathcal {U}_i(g,h)
                  \big],\label{eqd:dilaton}
\end{eqnarray}
where the $\tilde{c}_i$ and  $\chi_{\mu\nu}$ are defined in Eq.(\ref{eq4:chi}).

Now we introduce the metric ansatz for static black hole solution in $(d+1)$-dimensional space-time
\begin{eqnarray}\label{eqd:metric}
  ds^2 = -f(r) dt^2 +f^{-1}(r) dr^2+ r^2R^2(r) \sigma_{ij} dx^i dx^j,
\end{eqnarray}
where $\sigma_{ij} dx^i dx^j$\ is the line element for $(d-1)$-dimensional spherical Einstein space and volume $\omega_{d-1}$.
Considering the following reference metric
\begin{eqnarray}\label{eqd:ref metr}
  h_{\mu\nu}=diag(0,0,c_0^2 \sigma_{ij}),
\end{eqnarray}
the interaction potential in Eq.~(\ref{eq4:u1234}) changes into
\begin{eqnarray}\label{eqd:spec ui}
  \mathcal{U}_i &=&\Pi_i \frac{c_0^i}{R^i r^{i}}
\end{eqnarray}
with positive constant $c_0$   and  the notation $ \Pi_i=\prod_{j=0}^{i-1} (d-1-j)$.

In order to solve the system of Einstein equation (\ref{eqd:eist}),
we assume~\cite{Dehghani:2004sa,Sheykhi:2007wg}
\begin{eqnarray}\label{eqd:ansatz}
  R(r)=e^{\frac{2\alpha}{d-1} \varphi},
\end{eqnarray}
Then we can get the dilton field and the Liouville-type dilaton potential $V(\varphi)$ as the following form
\begin{eqnarray}
\varphi(r)&=&\frac{(d-1)\alpha}{2(1+\alpha^2)} \ln(\frac{\delta}{r})
\label{eqd:dilaton}\\
  V(\varphi)&=&\left\{
  \begin{array}{l}
    2\gamma_0 e^{2\xi_0 \varphi}+\sum_{i=1}^{4}2\gamma_{i}e^{2\xi_{i}\varphi}, \ \alpha \neq 1 \\
    2\lambda_0 \varphi e^{\frac{4 \varphi}{d-1}}+\sum_{i=1}^{4}2\lambda_{i} e^{2\zeta_{i}\varphi} ,\ \alpha = 1
  \end{array}\right.
  \label{eqd:dilaton:potential}
\end{eqnarray}
where the summation term is the general form associating with coupling between the dilaton and massive gravity field,
and $\lambda_0$,$\gamma_i$ and $\xi_i (i=0,1,...,4)$ are constants and satisfy
\begin{eqnarray}\label{eqd:parameters}
 \xi_0&=&\frac{2}{(d-1) \alpha}, \ \
 \xi_{i}=-\beta_{i}+ \frac {i}{(d-1)\alpha},\ \ \zeta_i=\frac{ i }{d-1}- \beta_i,\ \
  \gamma_0= \frac{\alpha ^2 (d-2) (d-1)}{2 \left(\alpha ^2-1\right) \delta ^2},\\ \nonumber
  \gamma_{i}&=&-\frac{m_0^2 c_0^{i}c_{i}\alpha \left(2 \alpha +(d-1) \beta
   _i\right) \delta ^{-i}\Pi_i}{2 \left(i-2 \alpha ^2+( 1-  d)\alpha \beta _i\right)},
\lambda_0=-\frac{2 (d-2)}{\delta ^2},\ \ \lambda_{i}=-\frac{ m_0^2 c_0^{i}c_{i}\Pi_i \delta ^{-i} ((d-1) \beta_i+2)}{2((1-d) \beta_i+i-2)}
\end{eqnarray}

 and then the solutions of Eqs. (\ref{eqd:eist})  in $(d+1)-$dimensional spacetime
can easily be calculated as
\begin{eqnarray}
  f(r)=\left\{
  \begin{array}{l}
  -m r^{\frac{\alpha
   ^2-d+2}{\alpha ^2+1}}
   -\frac{\left(\alpha
   ^2+1\right)^2 (d-2) \delta ^{-\frac{2 \alpha ^2}{\alpha
   ^2+1}} r^{\frac{2 \alpha ^2}{\alpha ^2+1}}}{\left(\alpha
   ^2-1\right) \left(\alpha ^2+d-2\right)}
    -\sum_{i=1}^4m_0^2 c_0^i c_i A_i r^{\frac{2 \alpha ^2+\alpha  (d-1) \beta _i-i+2}{\alpha
   ^2+1}}, \ \ \alpha\neq 1\\
  -m r^{\frac{3-d}{2}}-\frac{4 (d-2) r ((d-1) \log (\frac{r}{\delta})-d-1)}{(d-1)^2 \delta }
    -\sum_{i=1}^4m_0^2 c_0^i c_i B_i r^{\frac{ (d-1) \beta _i-i+4}{2}} ,\ \alpha = 1
  \end{array}\right.
      \label{eqd:metr all}
\end{eqnarray}
with
\begin{eqnarray}\label{eqd:nots}
A_i&=&\frac{ i\left(\alpha ^2+1\right)^2\delta ^{-\frac{\alpha  \left((d-1)
   \beta _i+\alpha  i\right)}{\alpha ^2+1}} \Pi_i}{(d-1) \left(\alpha ^2+\alpha  (d-1) \beta _i+d-i\right) \left(2 \alpha ^2+\alpha  (d-1) \beta
   _i-i\right)},\\ \nonumber
B_i&=& -\frac{4 i  \Pi _i \delta ^{\frac{1}{2} \left(-d \beta _i+\beta _i-i\right)}}{(d-1) \left((d-1) \beta _i-i+2\right) \left((d-1) \beta _i+d-i+1\right)}
\end{eqnarray}

Obviously, the solutions (\ref{eqd:metr all}) when taking $d=3$ equals to Eq.(\ref{eq4:metr fina}). According to the definition of ADM mass \cite{Abbott:1982},
the mass of dilatonic black hole reads as
\begin{equation}\label{eqd:adm mass}
  M=\frac{ (d-1) \omega_{d-1} \delta ^{\frac{\alpha ^2 (d-1)}{\alpha ^2+1}}m}{16 \pi  \left(\alpha ^2+1\right)},
\end{equation}
where $\omega_{d-1}$ represents the volume of hyper-surface
described by $h_{ij}dx^idx^j$.
Therefore,
using the  definition of horizon~$f(r_h)=0$,
the mass of black hole is written in terms of $r_h$ as
\begin{eqnarray}
M =\left\{
  \begin{array}{l}
   -\frac{\left(\alpha ^2+1\right) (d-2) (d-1)  \omega _{d-1} \delta ^{\frac{\alpha ^2 (d-3)}{\alpha ^2+1}} r_h^{\frac{\alpha ^2+d-2}{\alpha ^2+1}}}{16 \pi  \left(\alpha ^2-1\right) \left(\alpha ^2+d-2\right)}
   -\sum_{i=1}^{4} \frac{(d-1) m_0^2 A_i c_0^i c_i \omega _{d-1} \delta ^{\frac{\alpha ^2 (d-1)}{\alpha ^2+1}} r_h^{\frac{\alpha ^2+\alpha  (d-1) \beta _i+d-i}{\alpha ^2+1}}}{16 \pi  \left(\alpha ^2+1\right)},\ \alpha\neq 1,\\
   -\frac{(d-2)  \omega_{d-1}  \delta ^{\frac{d-3}{2}} r_h^{\frac{d-1}{2}} \left((d-1) \log (\frac{r_h}{\delta})-d-1\right)}{8 \pi  (d-1)}
   -\sum_{i=1}^{4}\frac{i m_0^2 \omega_{d-1}  c_0^i c_i \Pi _i \delta ^{\frac{(d-1) (1-\beta _i)-i}{2}} r_h^{\frac{(d-1) \beta _i+d-i+1}{2}}}{8 \pi  \left(i+2-(d-1) \beta _i\right) \left(i-2 -(d-1) (\beta _i+1)\right)},\ \alpha= 1.
  \end{array}\right.
  \label{eqd:mass}
\end{eqnarray}

By calculating the Hawking temperature and entropy of black hole  as
\begin{eqnarray}\label{eqd:temp}
T &=&\frac{1}{4 \pi}\frac{\partial f(r)}{\partial r}\big|_{r=r_h}\\ \nonumber
  &=&
  \left\{
  \begin{array}{l}
  -\frac{\left(\alpha ^2+1\right) (d-2)  \delta ^{-\frac{2 \alpha ^2}{\alpha ^2+1}} r_h^{\frac{\alpha ^2-1}{\alpha ^2+1}}}{4 \pi  \left(\alpha ^2-1\right)}
  -\sum_{i=1}^{4} \frac{m_0^2 A_i c_0^i c_i \left(\alpha ^2+\alpha  (d-1) \beta _i+d-i\right) r_h^{\frac{\alpha ^2+\alpha  (d-1) \beta _i-i+1}{\alpha ^2+1}}}{4 \pi  \left(\alpha ^2+1\right)},\ \alpha\neq 1,\\
  -\frac{(d-2) k \left(\log \left(\frac{r_h}{\delta}\right)-1\right)}{2 \pi  \delta }
  +\sum_{i=1}^{4} \frac{i m_0^2 c_0^i c_i \Pi _i \delta ^{\frac{(1-d) \beta _i-i}{2}} r_h^{\frac{(d-1) \beta _i-i+2}{2}}}{2 \pi  (d-1) \left((1-d ) \beta _i+i-2\right)},\ \alpha= 1.
  \end{array}\right.\\
  \label{eqd:entropy}
  S&=&
  \frac{1}{4} \omega _{d-1} \delta ^{\frac{\alpha ^2 (d-1)}{\alpha ^2+1}} r_h^{\frac{d-1}{\alpha ^2+1}}.
\end{eqnarray}
It's easily check that the first law of black hole thermodynamics still hold for dilatonic black hole in high dimensional spacetime
\begin{equation}\label{eqd: first law}
  d M = T dS
\end{equation}

\section{CONCLUSIONS and DISCUSSIONS}\label{5s}

Considering the non-minimal coupling between graviton and dilaton field, we discussed the massive Einstein-dilaton gravity.
According to the gravitational field and dilaton field equations by varying the action, we obtained the static spherically symmetric solutions of dilatonic black hole for $\alpha\neq 1$ and $\alpha= 1$ cases in four dimensional spacetime. Here the dilaton potential $V(\varphi)$ takes a so-called Liouville-type form both for $\alpha\neq 1$ and $\alpha= 1$, and the last two terms of potential are associated with the graviton terms.

Later, we have analyzed the singularity of the solution,
we give the Ricci and Kretschmann scalars,
which suggest that the horizons of black holes are just singularity of coordinate as it should be.
What is more, both of the scalars have the same the asymptotic behavior at the origin,
and one can obtain that there is a point of essential located at the origin,
and the asymptotic behavior of the solutions is not asymptotically flat but infinity in general when $r\to\infty$.
We also show that the  black hole solutions can
provide one horizon, two horizons (event and cosmological) , extreme (Nariai) and naked singularity black
holes for the suitably fixed parameters.
With dilaton field, the parameter $c_2$ affects the behavior of the metric function.

We further studied the mass, temperature and entropy of these dilatonic black holes and checked the first law of black hole thermodynamics. The analysis of the mass suggest that both for  $\alpha\neq 1$ and $\alpha= 1$, there could exists a maximum of the mass function of black hole horizon. For the temperature of these black holes,
and found that it has a minimum positive value $T_{min}$, distinguishing two sizes of black holes (a smaller one and a larger one) with the same temperature.
Moreover, the minimal points of the temperature function correspond
to the discontinuous  points of the heat capacity,
the domain of  smaller radii $r_h < r_{cri}$ of black holes lies instable and black holes with relatively large horizon radii $r_h > r_{cri}$ demonstrate stability,
which implies the occurrence of Hawking-Page phase transitions in these thermodynamic systems.
By investigating the Gibbs free energy, we have found the fact that large black holes always have a lower Gibbs free energy compared to small black holes confirming the arguments of their thermal stability.

Finally, we generalized these discussions to the high dimensional spacetime, and got the ($d+1$) dimensional solution of dilatonic black hole both for $\alpha\neq 1$ and $\alpha= 1$ in massive Einstein-dilaton gravity.
The corresponding thermodynamic quantities of black holes were also calculated, and we find the first law of black hole thermodynamics still maintains.

{\bf Acknowledgements}

We appreciate Yuxuan Peng for helpful discussion. B. Liu and Z. Y. Yang are supported by National Natural Science Foundation of China (Grants No. 12275213). D. C. Z is supported by National Natural Science Foundation of China (NSFC) (Grant No. 12365009) and Jiangxi Provincial Natural Science Foundation (No. 20232BAB201039). Q. Pan is supported by National Natural Science Foundation of China (Grant No. 12275079).

\appendix
\section*{APPENDIX: NEW FORM SOLUTION}

In this Appendix, we will present the static and spherically symmetric dilatonic black hole solutions with the traditional metric ansatz form in four dimensional spacetime.

We assume the new metric ansatz with
\begin{eqnarray}\label{eq4:metric1}
  ds^2 = -N^2(\tilde{r})f(\tilde{r}) dt^2 +f^{-1}(\tilde{r}) d\tilde{r}^2+ \tilde{r}^2 \left(d\theta^2+\sin^2\theta d\phi^2\right),
\end{eqnarray}
where $f(\tilde{r})$ and $N(\tilde{r})$ are functions of $\tilde{r}$.
Choosing the same reference metric in Eq.(\ref{eq4:ref metr}),
the interaction potential in Eq.~(\ref{eq4:u1234}) changes into
\begin{eqnarray}\label{u1:u2-1}
\mathcal{U}_1=\frac{2c_0}{ \tilde{r}},\ \ \
\mathcal{U}_2=\frac{2c_0^2}{ \tilde{r}^{2}},\ \ \
\mathcal{U}_3=\mathcal{U}_4=0,
\end{eqnarray}
and $\chi^{\mu}_{\nu}$ from Eq.(\ref{eq4:chi}) read as
\begin{eqnarray}\label{eq4:chi:1-1}
\chi^1_{~1} = \chi^2_{~2} = \frac{c_0 c_1 \tilde{r} e^{-2\beta_1 \phi }   + c_2 c_0^2 e^{-2\beta_2 \phi }}{\tilde{r} ^2 },\ \ \
\chi^3_{~3} = \chi^4_{~4} = \frac{c_1 c_0 e^{-2\beta_1 \phi }}{2 \tilde{r} }.
\end{eqnarray}

Then, the  corresponding
components of the Einstein equation Eq.(\ref{eq4:eist}) can be simplified to
\begin{flalign}
\tilde{G}^1_{~1}&=\frac{-1 + f(\tilde{r}) +\tilde{r} f'(\tilde{r})}{\tilde{r}^2}=-\frac{1}{2}V(\varphi)-f \varphi'^2+m_0^2 \chi^1_{~1}
\label{eq4:inseq:11-1},\\
 \tilde{G}^2_{~2}&=\frac{N'(\tilde{r}) [-1 + f(\tilde{r}) + \tilde{r} f'(\tilde{r})] + 2\tilde{r} f(\tilde{r}) N'(\tilde{r})}{\tilde{r}^2 N(\tilde{r})}=-\frac{1}{2}V(\varphi)+f \varphi'^2+m_0^2 \chi^2_{~2}
  \label{eq4:inseq:22-1},\\
\tilde{G}^3_{~3}&=\frac{1}{2\tilde{r} N(\tilde{r})} \left\{3 \tilde{r} f'(\tilde{r}) N'(\tilde{r}) + N(\tilde{r}) \left[2 f'(\tilde{r}) + \tilde{r} f''(\tilde{r})\right] + 2 f(\tilde{r}) \left[N'(\tilde{r}) + \tilde{r} N''(\tilde{r})\right]\right\} \nonumber \\
&=-\frac{1}{2}V(\varphi)-f \varphi'^2+m_0^2 \chi^3_{~3},
  \label{eq4:inseq:33-1}
\end{flalign}
where the prime $'$ denotes differentiation with respect to the radial coordinate $\tilde{r}$.

Considering the different forms between new line element \eqref{eq4:metric1} and traditional one \eqref{eq4:metric}, we assume the relationship of transformation of the two solutions
\begin{equation}
\tilde{r} =r R(r)= \delta^{\frac{\alpha^2}{\alpha^2 + 1}} \, r^{\frac{1}{\alpha^2 + 1}} ,
\qquad\text{or} \qquad r = \tilde{r}^{1 + \alpha^2} \delta^{-\alpha^2}.
\end{equation}
where $ R(r)$ is determined by Eqs.(\ref{eq4:Rr}) and (\ref{eq4:varphi}).
Based on Eqs.\eqref{eq4:chi:1-1}-\eqref{eq4:inseq:22-1},
 we obtain
 \begin{equation}\label{eq4:inseq:12-1}
\frac{N'(\tilde{r})}{\tilde{r} N(\tilde{r})}= \varphi'(\tilde{r})^2.
 \end{equation}
Therefore, the dilaton field can be written as
\begin{equation}
\varphi(\tilde{r}) = \alpha \ln\left( \delta/\tilde{r}\right)
\end{equation}
and the function $N(\tilde{r})$ reads as
\begin{equation}
N(\tilde{r}) = (1 + \alpha^2) \tilde{r}^{\alpha^2} \delta^{-\alpha^2}
\end{equation}
According to the similar calculations,
we found the differential  equation of dilaton potential $ V (\varphi (\tilde{r}))$ is same as Eq.\eqref{eq4:potential :dilation}
of $ V (\varphi (r))$.
So the total expansion of $ V (\varphi)$ remains consistent with Eq.\ref{eq4:di potential} in the previous text.
Finally, we can obtain the solution of metric function $f(\tilde{r})$ as
\begin{align}
f(\tilde{r}) =\left\{
 \begin{array}{l}
  -m \tilde{r}^{-\alpha ^2-1}+\frac{1}{1-\alpha ^4}-\frac{c_0 c_1 m_0^2 \tilde{r} \left(\frac{\delta }{\tilde{r}}\right)^{-2 \alpha  \beta _1}}{\left(\alpha ^2+2 \alpha  \beta _1+2\right) \left(2 \alpha ^2+2 \alpha  \beta _1-1\right)}
  -\frac{c_0^2 c_2 m_0^2 \left(\frac{\delta }{\tilde{r}}\right)^{-2 \alpha  \beta _2}}{\left(\alpha ^2+\alpha  \beta _2-1\right) \left(\alpha ^2+2 \alpha  \beta _2+1\right)}, \alpha\neq 1 \\
  1+\log (\delta/\tilde{r} )-\frac{m}{\tilde{r}^2}
  -\frac{c_0 c_1 m_0^2 \tilde{r} \left(\frac{\delta }{\tilde{r}}\right)^{-2 \beta _1}}{4 \beta _1^2+8 \beta _1+3}
  -\frac{c_0^2 c_2 m_0^2 \delta ^{-2 \beta _2} \tilde{r}^{2 \beta _2}}{2 \beta _2^2+2 \beta _2},\ \alpha=1
 \end{array}
\right.
\label{eq4n:solution}
\end{align}

As shown in Refs.~\cite{Babichev:2014ac}-\cite{Zou:2016sab}, the graviton mass does not significantly affect the form of the
entropy. According to the well-known entropy-area law,
the entropy as a pure geometrical quantity corresponding to the radius $\tilde{r}_h$ of  black hole can be obtained as
\begin{equation}\label{eq4:entropy-1}
S=\pi \tilde{r}^2=\pi  \delta ^{\frac{2 \alpha ^2}{\alpha ^2+1}} r_h^{\frac{2}{\alpha ^2+1}}.
\end{equation}

\end{document}